\documentclass{iaus}
\usepackage{graphicx}

\title[Abell 70] 
{A barium-rich binary central star in Abell 70}

\author[Henri M. J. Boffin et al.]   
{Henri M. J. Boffin$^1$,  B. Miszalski$^2$,  D. J. Frew$^3$,  A. Acker$^4$,\\
J. K\"oppen$^{4,5,6}$,  A. F. J. Moffat$^7$  \and  Q. A. Parker$^{3,8}$
}

\affiliation{
$^1$ESO, Santiago, Chile\\ email: {\tt hboffin@eso.org} \\$^2$South African Astronomical Observatory and SALT \\$^3$Macquarie University, Sydney, Australia\\ $^4$Observatoire astronomique de Strasbourg, France \\$^5$International Space University, Illkirch-Graffenstaden, France\\$^6$Universit\"at Kiel, Germany \\$^7$Universit\'e de Montr\'eal, Qu\'ebec, Canada \\$^8$Australian Astronomical Observatory}

\pubyear{2011}
\volume{283}  
\pagerange{000-000}
\jname{Planetary Nebulae, an Eye to the Future}
\begin{document}

\maketitle

\begin{abstract}
We have found the central star of Abell 70 (PN G038.1$-$25.4, hereafter A 70) to be a binary consisting of a G8 IV-V secondary and a hot white dwarf. The secondary shows enhanced Ba II and Sr II features, firmly classifying it as a barium star. The nebula is found to have Type-I chemical abundances with helium and nitrogen enrichment, which combined with future abundance studies of the central star, will establish A 70 as a unique laboratory for studying s-process AGB nucleosynthesis.
\end{abstract}

\firstsection 
\section{Barium stars}
Barium stars are peculiar red giants characterised by an overabundance of carbon and s-process heavy elements. The discovery that these stars were all binaries with white dwarf companions has led to the canonical model for the formation of these stars (\cite[Boffin \& Jorissen, 1988]{BoJo1988}). In this scenario the barium star was polluted -- in most cases while still on the main sequence -- by the wind of its companion that dredged up carbon and s-process elements during thermal pulses on the AGB. After ejecting its envelope as a PN, the AGB star then evolved into a white dwarf, while the contaminated star presents chemical anomalies: a barium star. Barium stars are important to better understand the s-process in AGB stars, mass transfer in binary systems and mixing processes. 

\firstsection
\section{Evidence for a barium star in A 70} The 2MASS survey recorded an unusually red $(J-H)=+0.70$ mag colour for the CSPN which suggested  the light is dominated by a cool companion, probably a subgiant or giant. Gemini GMOS and VLT FORS2 long-slit spectra confirmed the presence of the cool star, and indicated it is s-process enhanced, while GALEX observations reveal the presence of the hot white dwarf responsible for the nebula (\cite[Miszalski et al. 2011]{Brent2011e}, see Fig.~1). The radial velocities of the G8 IV-V star, as measured from the mean velocity of three Mg I absorption lines $\lambda$5167, 5172 and 5183, are consistent with the heliocentric nebula radial velocity, proving that the G8 IV-V star is physically connected to the nebula, as well as hinting at possible radial velocity variations due to orbital motion.
A 70 thus joins the growing club of the so-called A 35 type PN (A 35, LoTr 1, LoTr 5, and WeBo1; although A 35 may not be a true PN), that contain cool, rapidly rotating central stars exhibiting enhanced s-process abundances accompanied by very hot white dwarfs dominating at UV wavelengths (\cite[Bond et al. 2003]{Bond_2003}). These would therefore represent barium stars caught in a very transient stage, in which we can see both the polluted s-process rich cool star and the nebula ejected by the polluting star.

\begin{figure}[t]
\begin{center}
\includegraphics[width=5in]{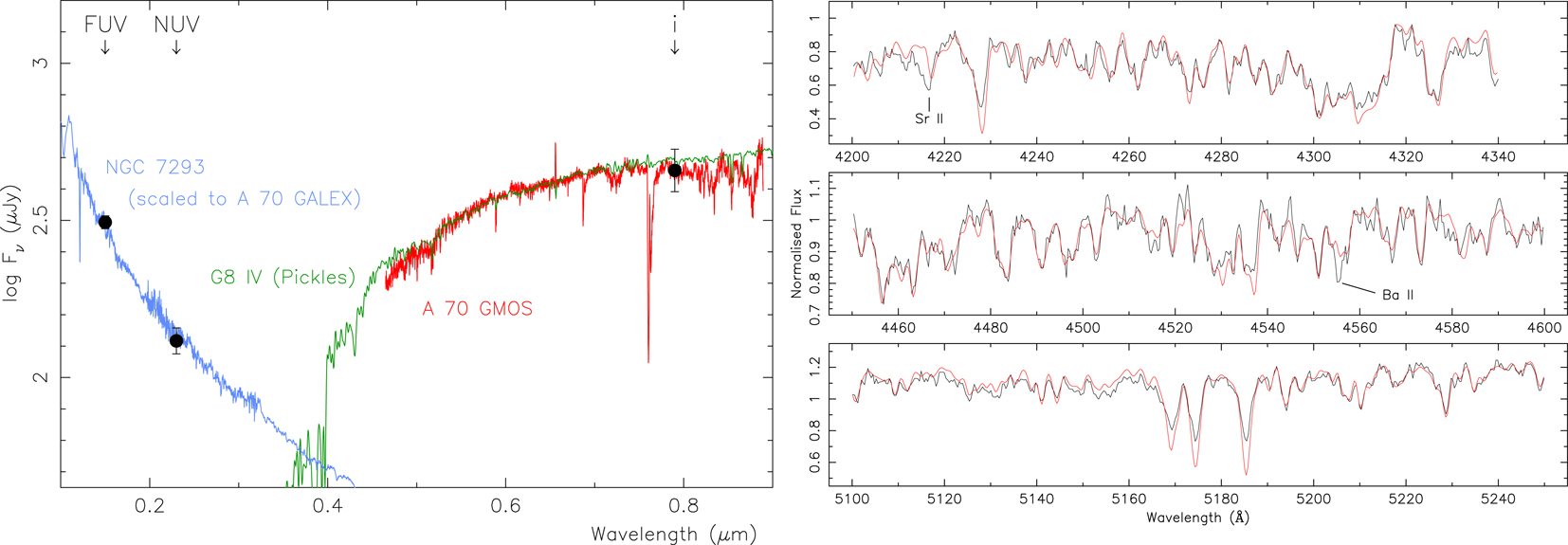} 
  
\caption{\small \emph{(left)} Spectral energy distribution of the binary nucleus of A 70. \emph{(right)} Normalised FORS2 spectrum (black) overlaid with the G8 IV-V comparison star HD24616. Note the absence of strong Sr II 4216 and Ba II 4554 in HD25616, whereas they are both enhanced in A 70.}
   \label{fig1}
\end{center}
\end{figure}

\firstsection
\section{Morphology and Chemical Abundances}
The apparent morphology of A 70 is that of a ring nebula. On closer inspection, the [O III] image shows a ridged appearance similar to Sp 1 (which is a bipolar nebula viewed close to pole-on; \cite[Bond \& Livio 1990]{BoLi1990}, \cite[Jones et al. 2011]{Jones2011}). Multiple knots of low-ionisation are also seen which are common in post-CE nebulae. It is unlikely however that the bipolar nebula of A 70 is the outcome of a CE interaction, as the barium star nature of A 70 most likely tells us that wind interaction in a long orbital period binary has happened and that this is the likely shaping mechanism for the bipolar nebula. Our deep spectra allowed us to measure accurate chemical abundances of the nebula, which was found to have a genuine Type-I composition with strong He and N enrichment.

\firstsection
\section{A Rosetta Stone for the late stages of stellar evolution}
Because of its evolutionary status -- having just left the main sequence and not yet gone through the first dredge-up that might dilute the accreted material -- the G8 IV-V secondary in A 70 presents us with a formidable opportunity to study several important aspects related to barium stars and planetary nebulae, in particular the s-process mechanism occurring in thermally-pulsing AGB stars, the common envelope phase, as well as mass transfer and wind accretion, all of which are still very far from being understood. We plan to perform a detailed study of A 70 in order to determine its orbital elements and derive the abundances of both low and high s-elements. 
 A 70 also offers a unique chance to detect Technetium (Tc) in a star that has been polluted by material coming from an AGB star, and, hence, proving the mass transfer scenario and the direct link between the cool star and the hot CSPN in A 70.

\end{document}